\documentclass[12pt]{iopart}
\usepackage{iopams}
\usepackage{graphicx}
\expandafter\let\csname equation*\endcsname\relax
\expandafter\let\csname endequation*\endcsname\relax
\usepackage{amsmath}
\usepackage{lineno}
\usepackage{color}
\usepackage{harvard}
\begin{document}
\review{The surprising Crab pulsar and its nebula: A review}
\author{R. B\"uhler$^1$, R. Blandford$^2$}
\address{$^1$Deutsches Elektronen Synchrotron DESY, D-15738 Zeuthen, Germany}
\address{$^2$Kavli Institute for Particle Astrophysics and Cosmology, Department of Physics and SLAC National Accelerator Laboratory, Stanford University, Stanford, CA 94305, USA}
\ead{rolf.buehler@desy.de, rdb@slac.stanford.edu}
\begin{abstract}
The Crab nebula and its pulsar (referred to together as ``Crab'') have
historically played a central role in astrophysics. True to their legacy,
several unique discoveries have been made recently. The Crab was found
to emit gamma-ray pulsations up to energies of 400~GeV, beyond
what was previously expected from pulsars. Strong gamma-ray flares, of durations of
a few days were discovered from within the nebula, while the source was
previously expected to be stable in flux on these time scales. Here we review these intriguing and suggestive developments. In this context we give an overview of the observational properties of the Crab and our current understanding of pulsars and their nebulae.

\end{abstract}

\maketitle
\tableofcontents
\clearpage
\section{Introduction}

On August 25, 1054 A.D., the Chinese astrologer Yang W\^elt reported ``the appearance of a guest star, above which some yellow-colored light was faintly seen''. He goes on to interpret the observations as showing that ``there is a person of great wisdom and virtue in the country. I beg that the Bureau of Historiography be given this message''\footnote{The term ``guest star'' was used by Chinese astronomers for new stars in the night sky, as novae or supernovae.} \cite{Peng-Yoke1972}. Yang W\^elt's message was heard and today we know that the new star which emerged in July 1054 A.D. on the night sky was the Crab supernova \cite{Breen1995}. Since then the remainder of this explosion, the Crab nebula and its pulsar, has been studied over the centuries. The source was rediscovered by an English amateur John Bevis who included it in a beautiful atlas which was never published because the publishers went bankrupt. He told his French colleague, Charles Messier about it. Messier was more interested in comets, to him the Crab Nebula was noise not signal. He included it in a famous catalog of celestial objects that could be confused with comets. The Crab is the first entry. The name of the nebula  is due to an Irishman, the third Earl of Rosse, who thought it looked like a Crab.

Following observations from J C Duncan in 1921, Edwin Hubble noted that, not only the Universe, but also the Crab nebula is expanding \cite{Hubble1928}. From the expansion velocity he correctly deduced that the nebula was the remainder of the 1054 A.D. star explosion, making the Crab the first historical supernova.  In one of the most celebrated and concise conjectures in the history of astronomy, Baade and Zwicky in 1934 ``advanced the view'' that in a supernova ``mass in bulk is annihilated'', ``cosmic rays are produced'' and that they ``represent a transition from ordinary stars into neutron stars'' \cite{Baade1934}. In 1942, Minkowski correctly associated one of the two central stars in the nebula with the explosion and in 1967 Pacini proposed that this star was a highly magnetized, spinning neutron star and that it powered the nebula \cite{PACINI1967}. The idea was put on a firmer footing by Gold who equated the rate of loss of rotational energy by the neutron star with the bolometric luminosity of the nebula \cite{Gold1968}. Almost 1000 years after the explosion of the progenitor star, the Crab is far from being quiet. At photon energies above $\gtrsim 30$~keV,  it is typically one of the brightest sources in the sky.


Due to its high luminosity $L \approx 1.3 \times 10^{38}$~erg~s$^{-1}$ \cite{Hester2008} and its proximity of $\sim 2$~kpc \cite{Trimble1973}, the Crab can be studied in great detail. It is therefore one of our prime laboratories to study non-thermal processes in the Universe. Many discoveries have been made in the Crab and then been seen in other non-thermal sources (including active galactic nuclei, gamma ray bursts and X-ray binaries) such as polarized synchrotron radiation or pulsed optical emission \cite{Shklovsky1953,Cocke1969}. True to this legacy  two remarkable discoveries have been made in the last years. Very high-energy (VHE $\gtrsim 100$~GeV) gamma-ray emission has been detected from the pulsar, and high-energy (HE $\gtrsim 100$~MeV) gamma-ray flares have been discovered from the nebula. These phenomena have not been observed in any other pulsar wind nebula to date. As we will discuss, they challenge and extend our understanding of these systems.

This article is structured as follows: first we will summarize the observational properties of the Crab and our current theoretical understanding of pulsar wind nebulae (section 2 \& 3). We will proceed to discuss the discovered gamma-ray pulsations and flares together with the ideas put forward so far to explain them (section 4 \& 5). Finally, we conclude the article with an outlook on future observational and theoretical developments. We would like to note that, given the wealth of work done on this source, it is not possible for us to be exhaustive. We will therefore give references to related review articles whenever possible. A more detailed description of the observational properties of the Crab can for instance be found in \citeasnoun{Hester2008}. A more detailed discussion of the theory of pulsar wind nebulae can be found in \citeasnoun{Kirk2009} and \citeasnoun{Arons2012}.

\section{Observational overview}

\begin{figure}[t]
\centering
\includegraphics[width=0.5\textwidth]{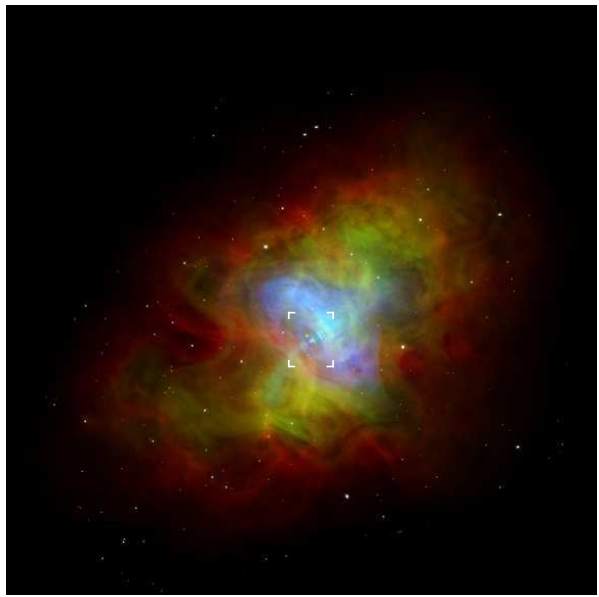}
\caption{A composite image of the Crab Nebula showing X-ray in blue, optical in green, and radio in red. The angular size of the image on the sky is 7.9~arc~minutes on each side \cite{Hester}. Credits: X-ray: NASA/CXC/ASU/J. Hester et al.; Optical: NASA/HST/ASU/J. Hester et al.; Radio: NRAO/AUI/NSF. The white box indicates the region shown in Fig. \ref{fig:dyn}.}
\label{fig:composite}
\end{figure}

The Crab nebula can be seen in a composite image in Fig. \ref{fig:composite}. The point source at its center is the Crab pulsar, the energy source of the system (see \citeasnoun{Rowan2004} and \citeasnoun{Harding2013a} for reviews on neutron stars and pulsars). Its energy is in its largest part emitted in a relativistic flow of magnetized plasma. This pulsar wind is thought to be predominantly composed of electron-positron pairs (referred to together as electrons in the following), although some ions might be present \cite{Gallant1994}. These pairs flow outwards and interact with the remaining ejecta of the progenitor star. As these particles spread out in the nebula, they loose energy due to synchrotron and inverse Compton radiation, creating the glowing pulsar wind nebula observed today. In the following we will describe the observations of the pulsar and the nebulae in more detail.

\subsection{The Crab pulsar}

\begin{figure}[t]
\centering
\includegraphics[width=0.7\textwidth]{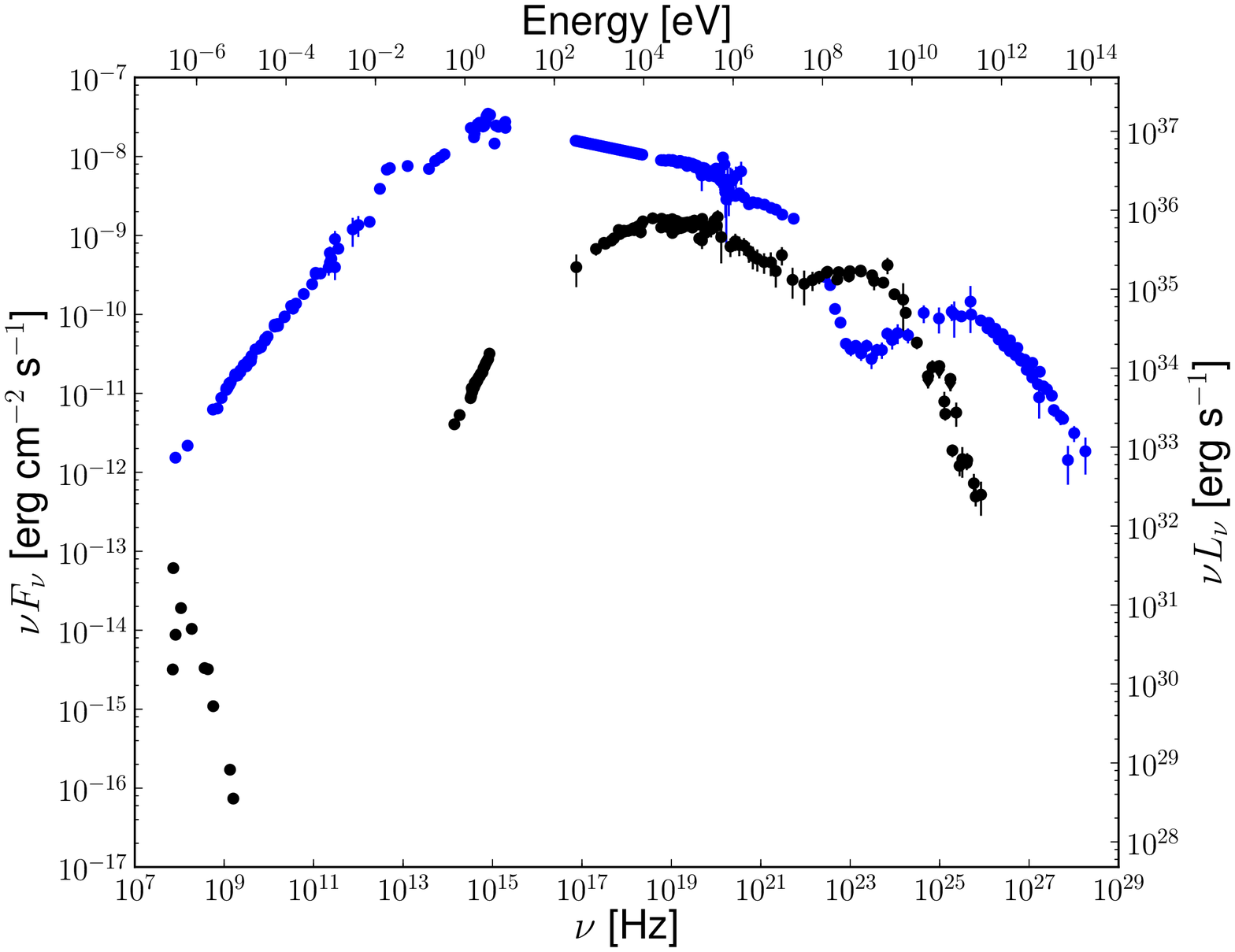}
\caption{Spectral energy distribution of the average emission of the Crab nebula (blue) and the phase averaged emission of the Crab pulsar (black). The data for the nebula were taken from \citeasnoun{Meyer2010} with the addition of the Fermi-LAT measurement reported in \citeasnoun{Buehler2012}. The pulsar spectrum is reproduced from \citeasnoun{Kuiper2001}. Additionally shown are infrared measurements reported in \citeasnoun{Sollerman2000} and \citeasnoun{Tziamtzis2009}, radio measurements referenced in \citeasnoun{Thompson1999} and gamma-ray measurements referenced in Fig. \ref{fig:pulsarsed}. Please note, that the low frequency radio data ($\lesssim 1$~GHz) comes from non-contemporaneous measurements, which are likely affected by time varying interstellar scintillation \cite{Rickett1990}. The luminosity shown on the right axis was calculated assuming a distance of 2 kpc.}
\label{fig:broadsed}
\end{figure}

The pulsar emits radiation across the electromagnetic spectrum with a period of $P_{\mathrm{Crab}} = 33.6$~ms, which is slowing down by $\dot P_{\mathrm{Crab}} = 4.2 \times 10^{-13}$ \cite{Manchester2005,Abdo2013}. The corresponding loss of rotational energy is $\dot E \approx 5 \times 10^{38}$~erg~s$^{-1}$, assuming a moment of inertia of the neutron star of $I \approx 10^{45} $~g~cm$^{-2}$. Only $\sim 1$\% of this energy is emitted in electromagnetic radiation. The radiation is thought to be collimated in beams of light, which sweep the field of view of the Earth, creating the observed pulsations. The phase-averaged spectral energy distribution (SED) is shown in Fig. \ref{fig:broadsed}. It is composed of three main components: the first one in the radio band, a second energetically dominant component peaking in the X-ray band and a third one emerging above energies of $\sim 100$~MeV.

\begin{figure}[t]
\centering
\includegraphics[width=0.4\textwidth]{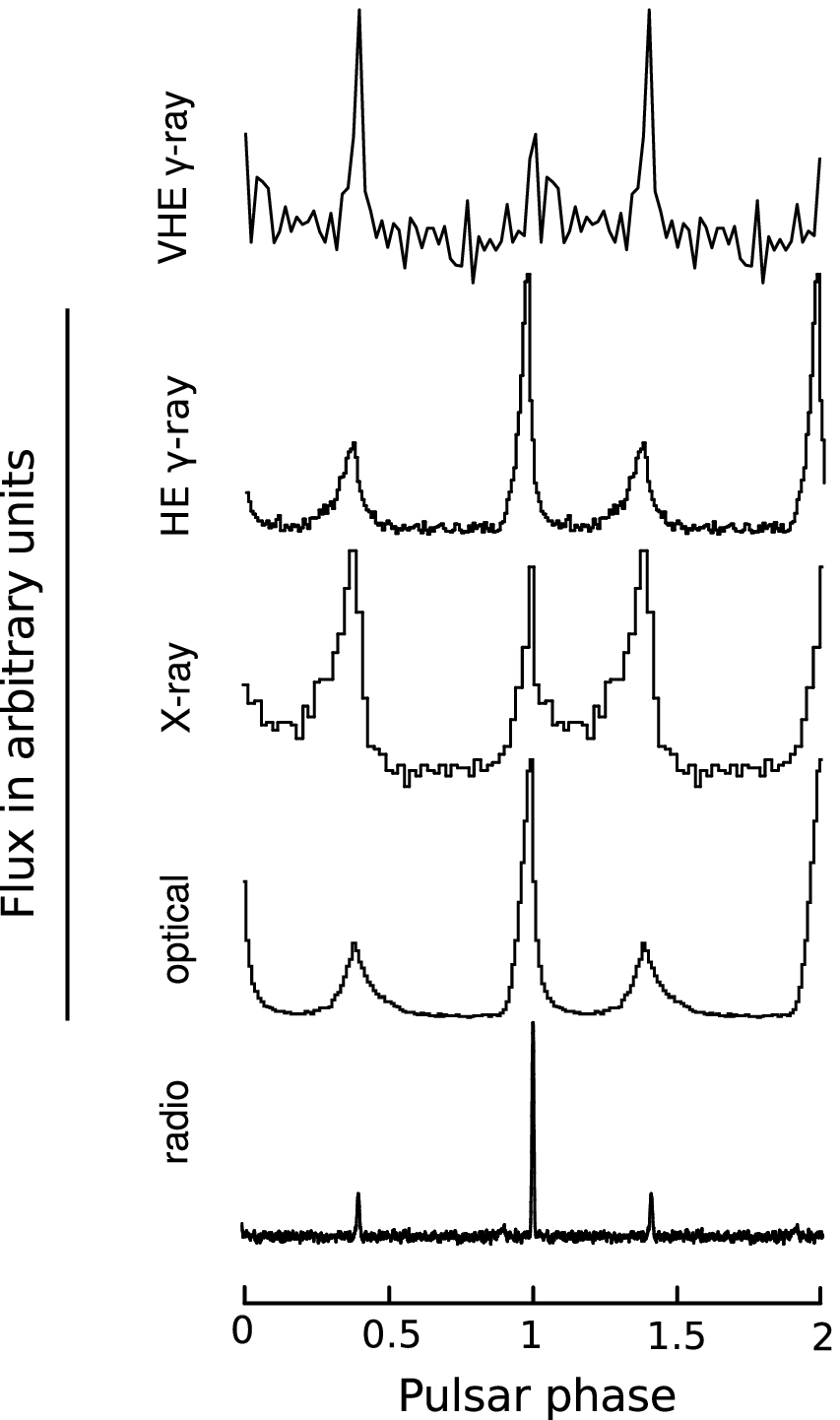}
\caption{Flux as a function of phase for radio (1.4 GHz), optical ($1.5-3.5$~eV), X-ray ($100-200$~keV), HE gamma-ray ($100-300$~MeV) and VHE ($50-400$~GeV) gamma-ray energies. The data is reproduced from \citeasnoun{Du2012}, with addition of the optical data from \citeasnoun{Oosterbroek2008} and VHE gamma-ray data from \citeasnoun{Aleksic2012}.}
\label{fig:phase}
\end{figure}

The phase profile of the emission is shown in Fig. \ref{fig:phase}. As for the majority of pulsars detected in gamma rays by the Fermi Large Area Telescope (LAT), the radiation as a function of phase $\phi$ is concentrated in two peaks \cite{Abdo2013}. The pulses arrive approximately simultaneously across the electromagnetic spectrum, only small shifts between energy bands are seen ($\Delta \phi \lesssim 0.01$; \citeasnoun{Oosterbroek2008} and \citeasnoun{Abdo2010}). The main pulse $P1$ is located at $\phi \approx 1.0$ and the inter-pulse $P2$ at $\phi \approx 0.4$. A faint precursor to $P1$ is detected at radio frequencies, and at frequencies above the optical bridge emission is found between $P1$ and $P2$. A peculiarity of the pulse profile of the Crab is that it is unusually irregular on short time scales. Giant radio pulses with a flux 1000 times the average are seen randomly during $P1$ and $P2$  \cite{Lundgren1995,Cordes2004,Popov2007}. Optical emission was seen to increase during such events \cite{Shearer2003,Strader2013}, but no such correlation has been found so far with higher energies \cite{Bilous2011,Bilous2012,Aliu2012}. Pulsar glitches, during which for a limited time the spin-down frequency increases by up to $\sim 10^{-5}$~Hz, are observed $\sim 1$ per year \cite{LyneA.G.1993,Espinoza2011,Wang2012}. 

The pulsar emission is found to be polarized. The position angle $PA$ of the linearly polarized component varies with pulse phase. The polarization properties in radio depend on frequency. Generally, the polarization degree and angle vary smoothly, with no abrupt changes during the main pulses \cite{Moffett1999}. The Crab is one of the few pulsars for which polarization has also been detected at optical wavelength \cite{Sowikowska2009,Moran2013}. The optical polarization angle swings from $PA \approx 40^{\circ}$ to $PA  \approx 170^{\circ}$ during P1, decreases again during the bridge emission and swings from $PA  \approx 90^{\circ}$ to $PA  \approx 180^{\circ}$ during P2.

\subsection{The Crab nebula}

\begin{figure}[t]
\centering
\includegraphics[width=0.95\textwidth]{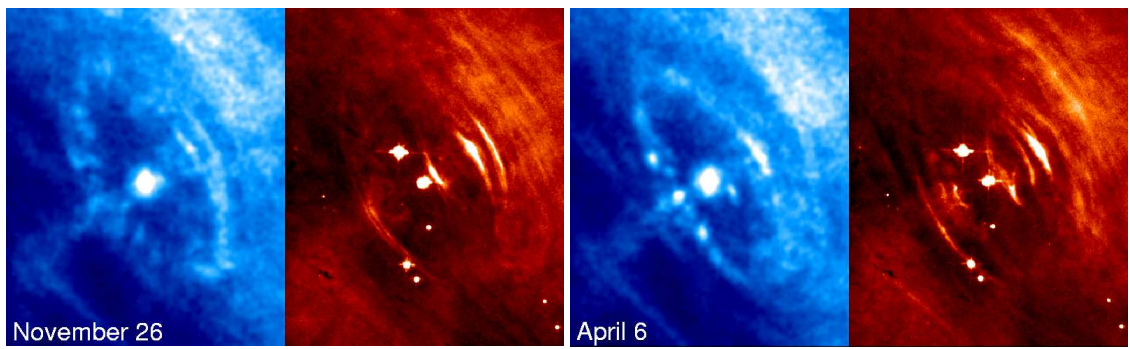}
\caption{The inner region of the Crab nebula is shown for the 26th of November 2000 (left panels) and the 6th of April 2001 (right panels). The images were taken during simultaneous observations of the Chandra X-ray Observatory and the Hubble Space Telescope \cite{Hester2002}. The X-ray image is shown in blue colors, the optical image in red colors. The angular size of the image on the sky is $0.7$~arc~minutes in the horizontal direction and $0.6$~arc~minutes in the vertical direction \cite{Hester}. Credits: X-ray: NASA/CXC/ASU/J. Hester et al.; Optical: NASA/HST/ASU/J. Hester et al..}
\label{fig:dyn}
\end{figure}

The appearance of the nebula in the sky is approximately ellipsoidal with a major axis of $\sim 7$~arc minutes and a minor axis of $\sim 4.6$~arc minutes. This corresponds to a projected length of $\sim 4.1$~pc and $\sim 2.7$~pc, respectively. As one observes the nebula at higher frequencies, a toroidal structure becomes increasingly apparent. Images of the inner region in X-rays and optical are shown in Fig. \ref{fig:dyn}. A torus surrounding the pulsar and a jet emerging perpendicular to it are apparent. It is striking that there is no bright emission in the region within $\sim 10$~arc~seconds of the pulsar \cite{Hester1995,Hester2002,Mori2004a,Temim2006}. The pulsar wind is apparently radiationless (or ``cold''), until interaction with the ambient medium happens. The first interaction is commonly thought to occur at the ``inner ring'' observed in the X-ray image \cite{Weisskopf2000}.

As can be seen in Fig. \ref{fig:dyn}, the inner nebula is a highly dynamical place. Thin arcs of increased emission, so called ``wisps'',  are observed to move outwards from the inner ring into the body of the nebula \cite{Scargle1969}. Wisps can be seen in radio, optical and X-rays, however their positions do not always coincide between frequencies. Typically, their inferred velocity is between 0.03~c and 0.5~c \cite{Hester2002,SchweizerThomas2013}, with indications for radio wisps to be slightly slower \cite{Bietenholz2001,Bietenholz2004}. Additionally, several other structures on scales of a few arc seconds are seen close to the inner ring and along the jet. Most prominently the ``Sprite'' is seen as a fuzzy region at the base of the jet in the optical images, and the ``inner knot''\footnote{There is some confusion with the nomenclature used in the literature. The ``inner knot'' is sometimes also referred to as ``knot 1'', and the ``Sprite'' sometimes referred to as ``Anvil''.} is detected only $\sim 0.6$~arc~second south east of the pulsar (the inner knot is too close to the pulsar to be visible in Fig. \ref{fig:dyn}, see e.g. \citeasnoun{Tziamtzis2009} or \citeasnoun{Moran2013}). Sprite, wisps and the inner knot are known to be variable down to time scales of a few hours \cite{Hester2002,Melatos2005}. 

Spectrally, the nebula shows a trend to softer spectra with increasing distance from the pulsar. This is interpreted as the radiative cooling of the highest energy electrons as they are convected and diffuse away from the inner nebula. Interestingly however, the spectrum in the inner region of the nebula is rather uniform. The torus, the jet and the small scale structures mentioned in the previous paragraph show no strong spectral variations among them from infrared to X-ray energies \cite{Veron-Cetty1993,Willingale2001,Temim2006,Mori2004a,Tziamtzis2009}. At radio frequencies the picture is more complex. As radio emission originates from a larger volume in the nebula, it is generally more difficult to disentangle line of sight effects. However, also at radio frequencies the spectrum is harder in the inner nebula \cite{Green2004a,Arendt2011}.

The emission from the nebula is found to be linearly polarized from radio to hard X-rays \cite{Wilson1974,Weisskopf1978,Bietenholz1991,Hester2008,Dean2008,Forot2008,Aumont2010}. Generally, the polarization angle $PA \approx 125^{\circ}$ is aligned with the symmetry axis of the nebula, indicating that the magnetic fields are predominantly azimuthal. A detailed study was recently performed in optical by \citeasnoun{Moran2013}. They find that the polarization of the wisps, the inner knot and the torus are all within a few degrees of the aforementioned $PA$ and have a high degree of polarization $PD \sim 50 \%$. While these structures are found to be variable in flux, their PA and PD show no significant variations.

The broad band SED of the nebula is shown in Fig. \ref{fig:broadsed}. One can see that the energy output of the nebula is a factor $\gtrsim 10$ larger than the one for the pulsar (in addition, the isotropic pulsar flux is likely even lower due to the beaming of its emission). In the nebula, the peak of the emission is located in the UV. The emission from radio to X-rays is known to be due to synchrotron emission  thanks to polarization measurements. At high energies ($\gtrsim 400$~MeV), a second emission component emerges due to inverse Compton emission of the same electrons. This component is energetically subordinate with respect to the synchrotron emission by a factor $\gtrsim 100$.

\section{Theory of pulsar wind nebulae}

\begin{figure}[t]
\centering
\includegraphics[width=0.7\textwidth]{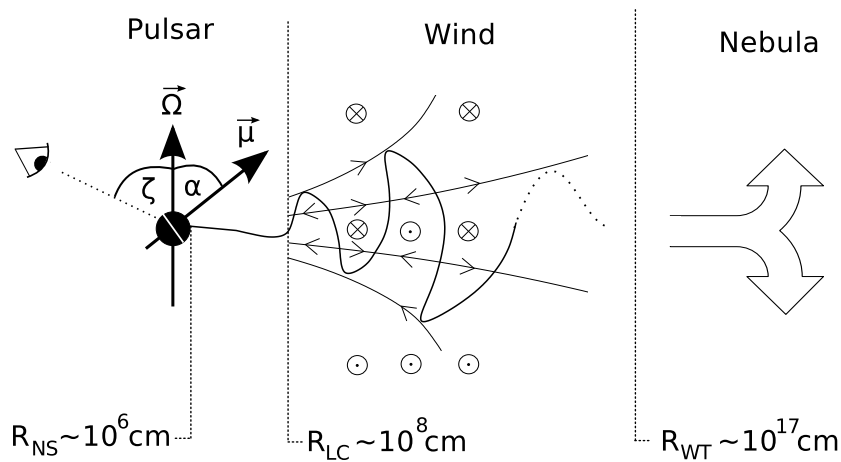}
\caption{Sketch of the components of a pulsar wind nebula discussed in the text. The equatorial current sheet (thick straight line) and the magnetic fields (thin dashed line and circles) are shown in the wind region (see text). The geometry of the system is illustrated, with $\zeta$ being the angle of the observer with respect to the rotation vector of the pulsar $\Omega$ and $\alpha$ being the inclination of the magnetic moment $\mu$ of the pulsar with respect to $\Omega$.  For the Crab $\zeta \approx 60^{\circ}$ was estimated assuming that $\Omega$ aligns with the symmetry axis of the nebula \cite{Ng2004}. The angle $\alpha$ is unknown and typically inferred to be $\approx 45^{\circ}$ from modeling of the pulsed emission \cite{Harding2008,Du2012}.}
\label{fig:sketch}
\end{figure}

To date $\sim 100$ pulsar wind nebulae have been detected, mostly at X-ray and TeV energies \cite{Kargaltsev2013}. The pulsar wind evolves inside of the supernova remnant of its progenitor star. Its time evolution is therefore complex, and depends on the properties of the progenitor and the environment in which it exploded. A review on pulsar wind nebulae evolution can be found in \citeasnoun{Gaensler2006}. In the following, we will focus on only on young systems ($\sim 1000$~years), which have not yet been significantly disrupted by the reverse shock of their supernova remnant, or the proper motion of their pulsar. The Crab is the best studied of such systems. 

In the common models of young pulsar wind nebula electrons and positrons are created in the magnetic fields of the pulsar and transported into the nebula. The electron density through most of the system is thereby expected to be large enough that magnetohydrodynamical conditions apply in good approximation. The pulsar wind is expected to carry most of the rotational energy lost by the pulsar into the nebula \cite{Rees1974}. The wind is flowing radiationless until its momentum flux is balanced by the ambient nebula pressure at a termination surface. At this surface, particles are randomized and begin to loose energy, predominantly through synchrotron radiation. Following these ideas the Crab can be subdivided in three regions shown in Fig. \ref{fig:sketch}: (i) The pulsar and its magnetosphere which extends out to the light cylinder radius $R_{\mathrm{LC}} = \frac{c P}{2\pi} \approx 1.4 \times 10^{8}$~cm (ii) The cold pulsar wind which is though to extend out to the inner ring. In the plane perpendicular to the symmetry axis of the nebula the latter is located at $R_{\mathrm{WT}}\approx 3 \times 10^{17}$~cm (iii) The synchrotron nebula, which extends from the inner ring into the outer nebula. We will discuss each of these regions in the following. 

\subsection{The pulsar magnetosphere}

A pulsar can be seen in first order as a rotating, perfectly conducting sphere, with a dipole magnetic field. The magnetic moment is generally not aligned with the rotational axis (the so called ``oblique rotator''). Compression of the magnetic field of the progenitor star onto a neutron star of $\sim 12$~km diameter results in magnetic fields of the order of $B \approx 3.8 \times 10^{12} (\frac{P}{P_{\mathrm{Crab}}})^{1/2} (\frac{\dot{P}}{\dot{P}_{\mathrm{Crab}}})^{1/2}$~G at the equatorial surface of the neutron star. The rotation of this field induces an electric potential of the order of $\Delta \Phi\approx 4 \times 10^{16} (\frac{\dot{P}}{\dot{P}_{\mathrm{Crab}}})^{1/2} (\frac{P}{P_{\mathrm{Crab}}})^{-3/2} $~V between the poles and the equator of the star. After obtaining vacuum solutions of the oblique rotator \cite{Deutsch1955}, it was realized that such strong potentials would remove particles from the neutron star surface and fill the magnetosphere with  plasma. The magnetosphere becomes charge separated to compensate the electric potential \cite{Goldreich1969}. This  makes the calculation of the magnetosphere properties, as magnetic field structure and electric densities and currents, intrinsically non linear, and dependent on the microphysics of the neutron star surface and the magnetospheric plasma. We therefore still do not have a fully self-consistent description of the magnetosphere of pulsars.

\begin{figure}[t]
\centering
\includegraphics[width=0.7\textwidth]{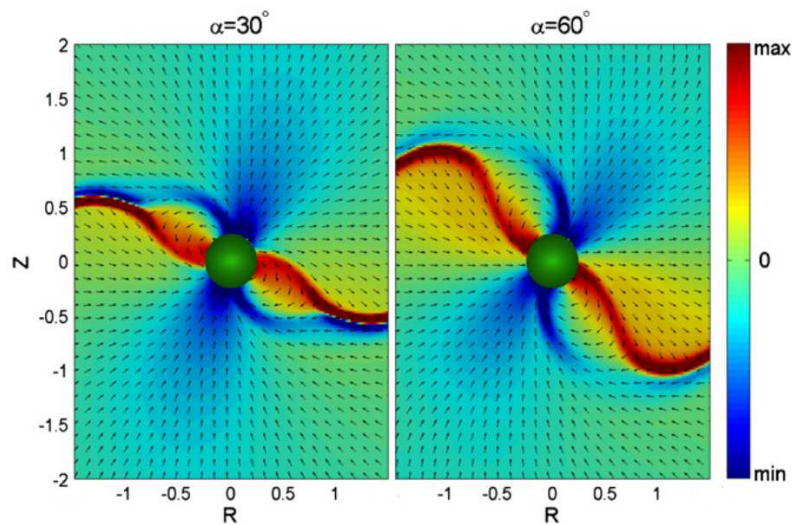}
\caption{Force free simulation of a pulsar magnetosphere reproduced from \citeasnoun{Bai2010}. Colors show of the charge density multiplied by the distance squared in the plane of the angular momentum $\Omega$ and magnetic moment $\mu$ of the pulsar. Results are shown for an inclination angle $\alpha = 30^{\circ}$ and $60^{\circ}$ between these two vectors. Arrows indicate the direction of the projected magnetic field. R and Z are in units of the light cylinder radius.}
\label{fig:mag}
\end{figure}

Over the past decade however, there have been significant progress in our understanding of the magnetosphere using the force free electrodynamics (FFE) approximation. The FFE approximation can be derived from magnetohydrodynamics (MHD) in the limit of inertia-free plasma, where the transport of energy and momentum is purely electromagnetical \cite{Komissarov2002,Blandford2002}. As in MHD, plasma is abundant and electric fields $E$ parallel to the magnetic fields $B$ are shortened out ($\vec{E} \cdot \vec{B} = 0$). There is no particle acceleration or resistivity. This  approach is supported by the fact that modeling of pulsar wind nebulae suggest that electrons are injected at high rates from the magnetosphere into the nebula $\dot N_{+-} > 10^5 \times \dot N_{\mathrm{GJ}}$, where $\dot N_{\mathrm{GJ}} =  7.6 \times 10^{33} (\frac{P}{P_{\mathrm{Crab}}})^{-1/2} (\frac{\dot{P}}{\dot{P}_{\mathrm{Crab}}})^{1/2}$~s$^{-1}$ is the ``Goldreich-Julian'' current \cite{Bucciantini2011}. Charge densities in the magnetosphere are therefore expected to significantly higher than the densities needed to compensate the electric potential due to pair creation within the magnetosphere (see \citeasnoun{Arons2012} and references therein). Further supporting the FFE approximation is that only $\lesssim 10 \%$ of the spin-down energy of pulsars is typically radiated in the form of pulsed emission, suggesting a low resistivity within the magnetosphere. Several simulations of magnetospheres in the FFE approximation have been made over the last years \cite{Spitkovsky2006,Kalapotharakos2009,Bai2010}. The charge density and magnetic field structure  in the plane of the angular moementum and magnetic moment vectors are shown in Fig. \ref{fig:mag}. The charge separation of the magnetosphere is apparent. A current sheet emerges in the equatorial plane, at the boundary between closed and open field lines\footnote{Field lines which return to the pulsar within the light cylinder are usually referred to as ``closed'', others as ``open''.} and crosses the light cylinder.

The FFE approximation cannot be exact, as there must be some resistivity in the magnetosphere. The latter depends on the properties of the co-rotating plasma, and has not been derived from first principles to date. However, ad-hoc introduction of resistivity into FFE and MHD simulations have been performed over the last years \cite{Li2012,Kalapotharakos2012a,Kalapotharakos2012b,Tchekhovskoy2013}. The basic magnetosphere properties, as e.g. the presence of the equatorial current sheet, are in agreement with the FFE approach. It was shown by \citeasnoun{Li2012} that the total spin-down luminosity and the potential drop over the open field lines transitions smoothly between the FFE approximation and the vacuum solution with increasing resistivity. The real magnetosphere of pulsars is therefore likely somewhere in between these two descriptions.

Neither MHD nor FFE simulations address the microphysics of the acceleration of particles within the magnetosphere. As currents flow out of the light cylinder on the open field lines, charge starved regions can emerge behind them (so called ``gaps''). An electric potential drop can build up in these regions until it is discharged by electron pair cascades and particles are accelerated. A polar gap is thought to emerge up to a distance of $\lesssim 10^{6}$~cm above the magnetic poles \cite{Sturrock1971,Ruderman1975,Holloway1975,Daugherty1982,Baring2004}. Gaps location have also been proposed out in the magnetosphere, along the last open field lines (slot gap \cite{Arons1979,Muslimov2004}; annular gap \cite{Qiao2004,Du2012}), reaching out to the light cylinder (outer gap \cite{Cheng1976,Cheng1986,Romani1995,Cheng2004}). Alternatively, particles might also be accelerated in reconnecting magnetic fields at the equatorial current sheet \cite{Li2012,Tchekhovskoy2013}. In either case, the particles loose their energy after leaving the acceleration region due to interaction with the magnetic and photon fields of the magnetosphere, and secondary particle cascades might be induced. As the direction of the particles is bound closely to the magnetic field lines, co-rotating cones of light are emitted around both magnetic poles. As one or both of them pass our line of sight, the observed electromagnetic pulsations are produced.

\subsection{The cold pulsar wind}

It was shown at the early days of pulsar studies, that the magnetic field lines become asymptotically radial beyond the light cylinder in the FFE approximation \cite{Ingraham1973,Michel1974}. This justified studying the fields and flows outside of the light cylinder in the ``split-monopole'' approximation, where the field lines from the pulsar are thought to be a monopole which inverses its field direction at the equator. An analytic solution to the obliquely rotating split-monopole was found by \citeasnoun{Bogovalov1999}. Generally, plasma that flows outside of the magnetosphere on the open field lines begins to trail the pulsar rotation, creating a helical pattern. A current sheet emerges which undulates within an angle $\alpha$ around the equatorial plane (the so called ``striped wind''). The magnetic field lines become predominantly toroidal, and reverse their direction at the current sheet (see Fig. \ref{fig:sketch}). Qualitative agreement with the split-monopole solutions has recently been found in FFE and MHD simulations out to a distance of $\lesssim 10 R_{\mathrm{LC}}$ \cite{Kalapotharakos2012,Tchekhovskoy2013}.

The Poynting flux per solid angle of the split monopole solution scales as $P \propto sin^2\theta$, where $\theta$ is the angle with respect to the rotation axis of the pulsar. As most of the energy is emitted around the equatorial plane, the striped wind plays an important role in the dynamics of the nebula.  The wind is thought to be magnetically dominated at its launch, with $\sigma >> 1$, where $\sigma$ is the ratio of the magnetic to the kinetic energy of the flow. However, as we will discuss in the next section a low-sigma flow of $\sigma \sim 0.003$ was inferred at the termination of the wind from 1D MHD modeling of the synchrotron nebula. The emerging question how magnetic energy is transferred to particle energy in the flow is referred to as the ``$\sigma$-problem''. In principle, at low latitudes the alternating magnetic fields of the striped wind make it susceptible to magnetic reconnection due to instabilities in the current flow \cite{Coroniti1990}. However, it was realized that due to the relativistic motion of the wind, the time scales of these processes are likely too long \cite{Lyubarsky2001,Kirk2003}. Recently, a new paradigm is emerging were magnetic energy is thought to be dissipated predominantly after the wind termination, as we will discuss in the next section.

\subsection{The synchrotron nebula}

\begin{figure}[t]
\centering
\includegraphics[width=0.95\textwidth]{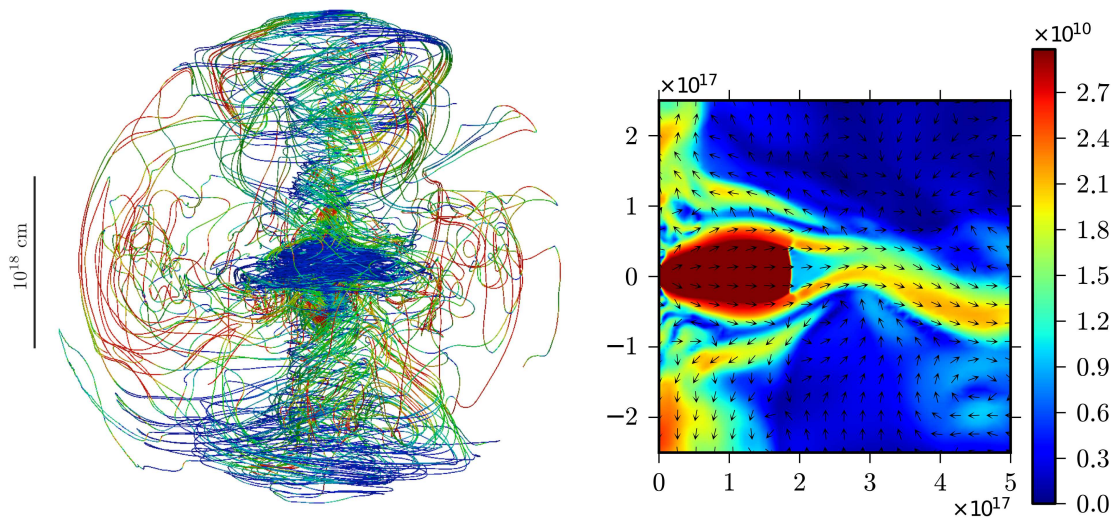}
\caption{\textbf{Left:} sample magnetic field lines of a three dimensional MHD model of the Crab nebula with $\sigma = 3$, reproduced from \citeasnoun{Porth2013}. Colors indicate the dominating field component, blue for toroidal and red for poloidal. \textbf{Right:} velocity magnitude and direction of the plasma for the same simulation. Reproduced by permission of Oxford University Press.}
\label{fig:porth}
\end{figure}

As the cold pulsar wind can not be observed before it termination, its properties can only be inferred indirectly by observing the synchrotron nebula. Spherical symmetric MHD solution downstream of the termination surface were derived by \citeasnoun{Kennel1984}, predicting the magnetic field distribution in the nebula. The key parameter of this model is the magnetization of the wind just before its termination. In particular the distance of the termination surface from the pulsar in the Crab could only be reproduced if the magnetization at this point is low $\sigma \sim 0.003$. The magnetic field strength is then expected to increase behind the shock and have values between 100-300 $\mu$G in the synchrotron nebula \cite{DeJager1992,AtoyanA.M.1996,Hillas1998,Meyer2010,Martin2012,Torres2013}. This field is lower than the one expected in the case of equipartition between particle energy and magnetic fields, indicating that $\sim 1/30$ of the internal energy in the nebula is in magnetic form \cite{Hillas1998}. Similar magnetic field values are inferred by interpreting the hardening of the integrated spectrum of the nebula between the radio and the optical band due to electron cooling \cite{Marsden1984}.

It is typically assume that particles are re-accelerated at the wind termination surface. However, it is unlikely that diffusive shock acceleration of particles takes place, as the emerging shock is expected to be relativistic and oblique \cite{Sironi2011}. Instead, the turbulence downstream of the shock might trigger magnetic reconnection, which can lead to non-thermal particle acceleration \cite{Lyubarsky2003}. Particle in cell (PIC) simulations show that, independent of the wind magnetization, the energy of the alternating field components of the striped wind is dissipated into kinetic energy downstream of the wind termination \cite{Sironi2011a}. Assuming that electrons are accelerated into a power-law spectrum at the termination surface one dimensional MHD models reproduces the spectral energy distribution of the nebula qualitatively, with the exception of the radio band \cite{AtoyanA.M.1996,Volpi2008}. As the life time of radio emitting electrons is much longer than the age of the nebula, these particles might have been injected in the past. Radio emission might therefore come from a separate electron population, which could have been injected during  the high spin-down phase of the pulsar following the super novae explosion \cite{AtoyanA.M.1996}. Alternatively, radio emitting electrons could be continuously accelerated throughout the nebula due to magnetic reconnection in MHD turbulence \cite{Nodes2004}. Magnetic reconnection downstream of the wind termination can also reproduce the spectrum of radio emitting electrons, if electrons are injected at a rate of $\dot N_{+-} \sim 10^8 \times \dot N_{\mathrm{GJ}}$ into the pulsar wind and the magnetization is high with $\sigma>30$ \cite{Sironi2011a}. The implications of high electron multiplicities therefore deserve further study.

While the one dimensional models of the Crab laid out our general picture, it was clear form the beginning that the asymmetry of the nebula needs to be taken into account for a more quantitative description of the nebula. The dynamics of the flow at and beyond the wind termination have therefore been studied with multidimensional MHD simulations over the past decade \cite{Komissarov2004,Bucciantini2006,Porth2013}. The properties of the pulsar wind at its termination and the particle energy spectrum after the wind termination are free parameters of these simulations. Tracing the maximum particle energy in the flow makes it possible to compute synchrotron, inverse Compton and polarization maps of the nebula, which can be compared to observations \cite{delZanna2006,Volpi2008,Volpi2009}.  Axially symmetric models with an appropriately chosen azimuthal dependence of the wind qualitatively reproduce the Crab morphology: a torus emerges, and the flow is bend backwards due to the ``hoop-stress'' in the downstream, creating a jet (see Fig. \ref{fig:porth}). They explain the emergence of wisps and their dynamics. Their increased brightness is thought to be predominately due to Doppler beaming towards our line of sight \cite{Camus2009}. The simulations predicted variability of a few percent per year in the overall X-ray emission of the nebula \cite{Volpi2008}, which has recently been discovered \cite{Wilson-Hodge2011,Kouzu2013}. This is remarkable in high-energy astrophysics, where predictions and measurements are seldom  done on this level of accuracy. 

As the early 1D MHD solution, the 2D simulations show that a low-sigma wind with $\sigma\sim0.02$ matches the observed morphology, and the location the shock at the observed $R_{\mathrm{WT}}$. Recently, the first 3D simulations were performed \cite{Mizuno2011,Porth2013,Porth2013b}. Interestingly, these simulations show that the morphology of the Crab can also be obtained at high magnetization ($\sigma>1$). As was pointed out by \citeasnoun{Begelman1998}, 1D and 2D MHD solutions prevent the growth of turbulence due to current instabilities. Due to this turbulence, magnetic fields are randomized  and remain predominantly axial only close to the wind termination (see Fig. \ref{fig:porth}), and pressure and magnetic field gradients in the nebula are reduced when compared to 1D or 2D models. The increased turbulence induces magnetic dissipation in the downstream, alleviating the $\sigma$-problem  \cite{Lyutikov2010}.  While it might be too early for a final verdict, it appears that the origin of the latter lay in the simplifications made in 1D and 2D models.

The success of the MHD simulations in modeling many of the nebula's properties shows that the general picture we have outlined is likely a good approximation of reality. However, several observational findings can not be explained or have not been studied yet. A quantitative comparison of the morphology obtained in the simulations to the observed one has not been performed to date. In particular, the inner ring does not seem to be found consistently in the simulations. At odds with the expectation from Doppler boosting, the Chandra observations show that the inner ring has similar brightness at the front and the receding side. We also do not  have an explanation why the ring appears to be composed of a series of knots, which are highly variable in time \cite{Weisskopf2000,Porth2013b}. The observation that the wisps at different wavebands are generally not aligned has not been studied in the simulations. If Doppler beaming is predominantly responsible for their enhanced brightness, one would naively expect radio, optical and X-ray wisps to be co-spatial. In addition, 2D MHD simulations with low magnetization overestimate the observed inverse Compton emission \cite{Volpi2008}. It will be interesting to see if this discrepancy is reduced in 3D models of high-magnetization.

\section{Pulsed very high-energy gamma-ray emission}

After having discussed the observational properties of the Crab and our theoretical understanding of pulsar wind nebulae, we will turn to the recent discoveries and their implications in the following. We begin with the detection of pulsed emission in VHE gamma rays. The emission of pulsars is typically expected to fall close to exponentially $\nu F_{\nu} \propto exp \left( {\frac{E}{\epsilon_{\mathrm{cut}}}} \right)^b$ above a cutoff energy $\epsilon_{\mathrm{cut}}$. This cutoff is thought to be inherent to the emission mechanism, as we will discuss later. Indeed, all spectra of the $\sim 150$ pulsars detected to date at HE-gamma-rays are compatible with this expectation \cite{Abdo2013}. Surprisingly, a significantly slower spectral decay is observed for the Crab pulsar.

\subsection{Observational results}

\begin{figure}[t]
\centering
\includegraphics[width=0.7\textwidth]{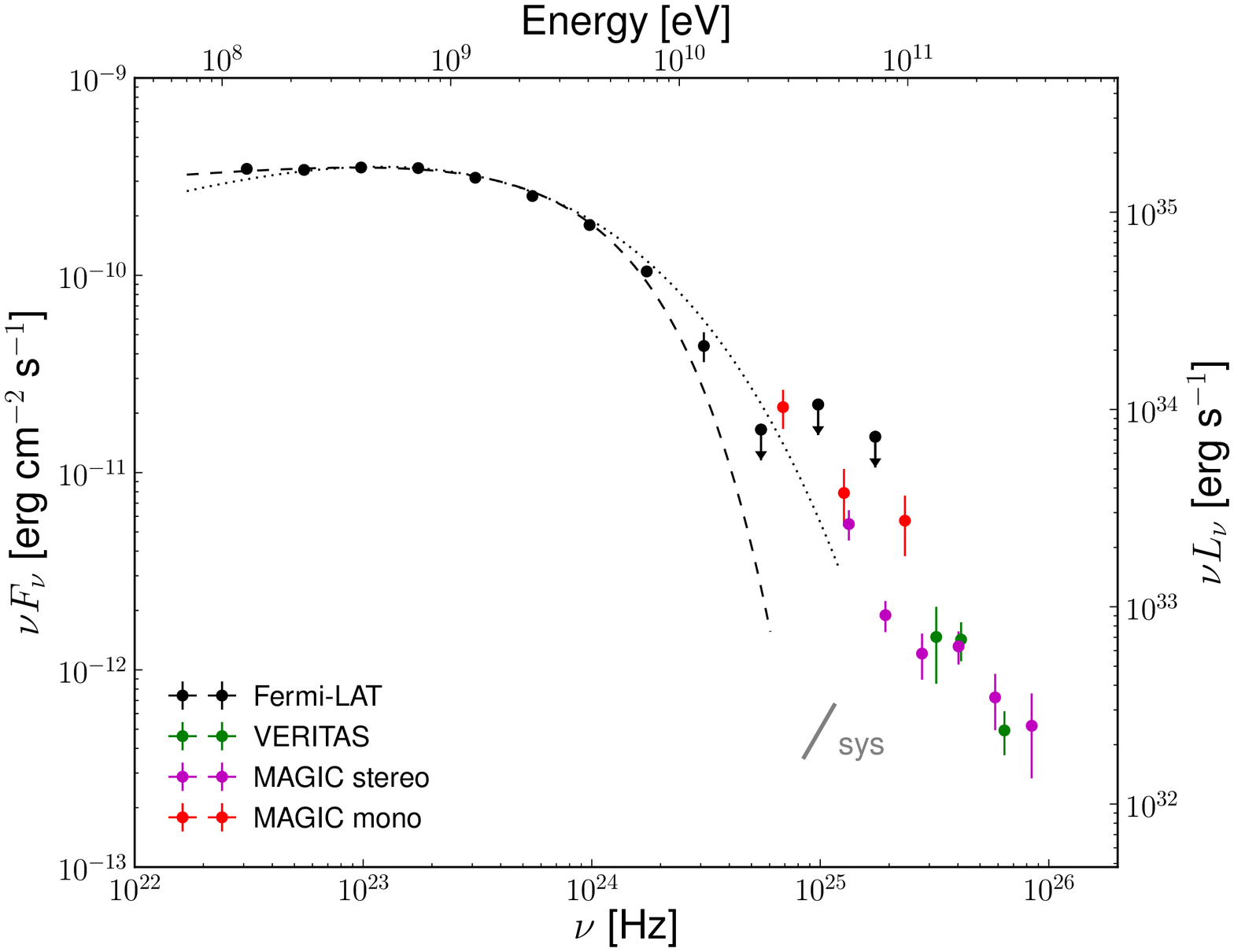}
\caption{Phase averaged spectral energy distribution of the Crab pulsar in gamma-rays. The spectra are reproduced from \citeasnoun{Abdo2013}, \citeasnoun{Aliu2011} , \citeasnoun{Aleksic2011} and \citeasnoun{Aleksic2012}. The effect of systematic shifts of $\pm 15 \%$ on the normalization of the energy scale is shown by the gray line. These are the typical systematic errors of VHE gamma-ray instruments. Best fit parametrizations of the Fermi-LAT data of  a power-law function with an exponential (dashed) and sub-exponential (dotted) cutoff are shown (see text).}
\label{fig:pulsarsed}
\end{figure}

The high-energy end of the SED of the Crab pulsar is shown in Fig. \ref{fig:pulsarsed}. The best fit parametrizations of the Fermi-LAT measurement by a power-law function with an exponential cutoff ($b=1$) and a sub-exponential cutoff ($b<1$) are also displayed. The latter parametrization is preferred statistically at a $\gtrsim 3 \sigma$ confidence level, indicating a slower than exponential fall of the spectrum. This is not unusual for LAT detected pulsars \cite{Abdo2013}. However, unexpectedly VHE gamma-ray measurements showed an enhanced emission above an extrapolation of both of these parametrizations \cite{Aliu2008,Aliu2011,Aleksic2011,Aleksic2012} \footnote{Statistically marginal experimental discrepancies can be seen between different measurements in Fig. \ref{fig:pulsarsed}. In particular, the MAGIC mono data have larger fluxes compared to the other measurements. However, within systematic errors these differences are not significant.}. Considering systematic errors, the Fermi-LAT and VHE-gamma-ray data can be described by one power-law function above $E \approx 4$~GeV \cite{Aliu2011}. No indication of a spectral cutoff has been found up to photon energies of $\sim400$~GeV.

The phase distribution of the VHE gamma-ray events is shown in Fig. \ref{fig:phase}. The peaks of the gamma-ray pulses coincide with the radio ones. Compared to the HE-gamma-ray pulsations, the VHE ones are narrower. This narrowing of the pulse width with increasing energy can also be seen in the HE and VHE data independently \cite{Abdo2010,Aleksic2012}. Interestingly, a hardening of the spectrum of $P2$ with respect to $P1$ is observed in similar form in two components of the SED: the relative flux of $P2$ increases with respect to $P1$ from radio to hard X-rays, and from HE gamma-rays to the VHE gamma-rays.

\subsection{The origin of the gamma-ray pulsations?}

While radio emission from pulsars is usually attributed to coherent emission close to the polar gap, at higher frequencies the emission has to come from outer parts of the atmosphere. In particular the gamma-ray emission is expected to come from distance of $0.1-1 R_{\mathrm{LC}}$ \cite{Romani1995,Qiao2004,Muslimov2004}. This is inferred from the phasograms and spectra of the population of gamma-ray detected pulsars \cite{Abdo2013}. For the Crab pulsar, this expectation is confirmed in a model-independent way by the detection of emission beyond $100$~GeV. To avoid absorption of gamma rays due to pair production on the pulsars magnetic fields, the photons have to be emitted at a distance $\sim 10^7$~cm~$\approx 0.1 R_{\mathrm{LC}}$ from the neutron star, a factor $\gtrsim 10$ above the expected location of the polar gap (Eq. 1 in \citeasnoun{Baring2004}).

The emission mechanism at the high-energy end of the SED of pulsars is expected to be dominated by curvature radiation of particles streaming along magnetic field lines.  However, the high energy of the pulsations makes this unlikely for the Crab. Radiative losses due to curvature radiation would produce a cutoff in the spectrum above gamma-ray energies of  $\epsilon_{\mathrm{cut}}=150$~GeV~$(E/B)^{3/4} \sqrt{R / R_{\mathrm{LC}}}$, where $E$ is the accelerating electric field \cite{Aliu2011}. The latter is typically assumed to be $E \lesssim 0.1 B$ in pulsar gaps. The absence of a cutoff in the spectrum up to energies $>100$~GeV therefore implies $R > R_{\mathrm{LC}}$. As particles stream along magnetic field lines only within the light cylinder, curvature radiation is likely not responsible for the VHE pulsations.

It was proposed by \citeasnoun{Lyutikov2012} that the emission at the high-energy end of the SED is dominated by Inverse Compton scattering.  Indeed, the pulsed emission can be explained by up-scattering of photons in particle cascades induced by outer gaps \cite{Aleksic2011,Aleksic2012} or annular gaps \cite{Du2012}. Alternatively, Inverse Compton emission can also occur in the striped wind \cite{Bogovalov2000,Kirk2002}.  As proposed by \citeasnoun{Aharonian2012} and \citeasnoun{Petri2012}, the VHE emission may result from the up-scatter of pulsed X-ray photons. If this is the case, the VHE observations are the first direct observational signature of the cold pulsar wind. In order to fit the observed pulse profile and spectrum of the VHE emission, the wind needs to be abruptly accelerated at $20-50 R_{\mathrm{LC}}$. However, while the gamma-ray emission can be explained in this picture, the observed narrowing of the VHE pulses with increasing energy would not be expected.

\section{Gamma-ray flux variations from the nebula}

Flux variability on short time scales from structures within the Crab has been known for almost 100 years \cite{Lampland1921}. However, the flux integrated over the volume of the nebula is varying only slowly over time: evidence for $\sim 1 \%$ per year variations have been found in radio, optical and X-ray wavelength \cite{Vinyaikin2007,Smith2003,Wilson-Hodge2011}. As this is small compared to the systematic errors of most high energy telescopes, the Crab was often used to cross-calibrate instruments, particularly in the X-ray and TeV domains. Stronger variability was generally expected to be found at the high end of the synchrotron component ($\sim 100$~MeV), as this emission is expected to be emitted by particles with $\sim 1$~PeV energies in a $\sim 0.1$ mG magnetic field. The cooling time scales of these particles is $\sim 1$~year, and indeed  indications for variability of $\sim 25 \%$ over two years were found in this frequency range \cite{Much1995,deJager1996}. Nevertheless, it came as a huge surprise when the Fermi-LAT and AGILE satellites detected strong gamma-ray flares, with increases of the unpulsed flux by a factor $\sim 30$ above 100 MeV on time scales down to $\sim 6$~hours \cite{Abdo2011,Tavani2011,Balbo2011,Ackermann2013}. We will discuss the observational results on the HE gamma-ray variability of the nebula in the following and summarize the theoretical conclusions drawn so far afterwards.

\subsection{Observational results}

The gamma-ray flares are observed in the energy range between the synchrotron and inverse Compton component of the spectral energy distribution. The average  photon flux of the synchrotron component is $F_{\mathrm{ns,100}} = (6.1 \pm 0.2 ) \times 10^{-7}$~cm$^{-2}$~s$^{-1}$ and the one of the inverse Compton component is $F_{\mathrm{ns,100}} = (1.1 \pm 0.1 ) \times 10^{-7}$~cm$^{-2}$~s$^{-1}$ above 100 MeV. For comparison, photon flux of the pulsar above this energy is $F_{\mathrm{p,100}} = (20.4 \pm 0.1 ) \times 10^{-7}$~cm$^{-2}$~s$^{-1}$. The emission of the inverse Compton component of the nebula and the pulsar flux are found to be constant in time within measurement accuracies. However, in the synchrotron nebula is variable on all time scales which can be resolved. The power density spectrum as a function of frequency of the flux variations is compatible with a power-law $PDS \propto \nu^{-0.9}$ \cite{Buehler2012}. The flux can remain below the detection threshold of the Fermi-LAT for a month, with flux upper limits well below the average flux value \cite{Abdo2010}. On the other hand, the flux can rapidly increase within a few hours during flares. Whether or not the flares are distinct events or the high-energy end of a continuous spectrum in variability is unclear. Time scales down to $\approx 10$~hours can be resolved in the PDS with no sign of a break in the spectrum. However, the flares are far outlayers in the flux distribution and the spectrum during the brightest flares clearly shows the emergence of a new spectral component in the SED.

\begin{figure}[t]
\centering
\includegraphics[width=0.7\textwidth]{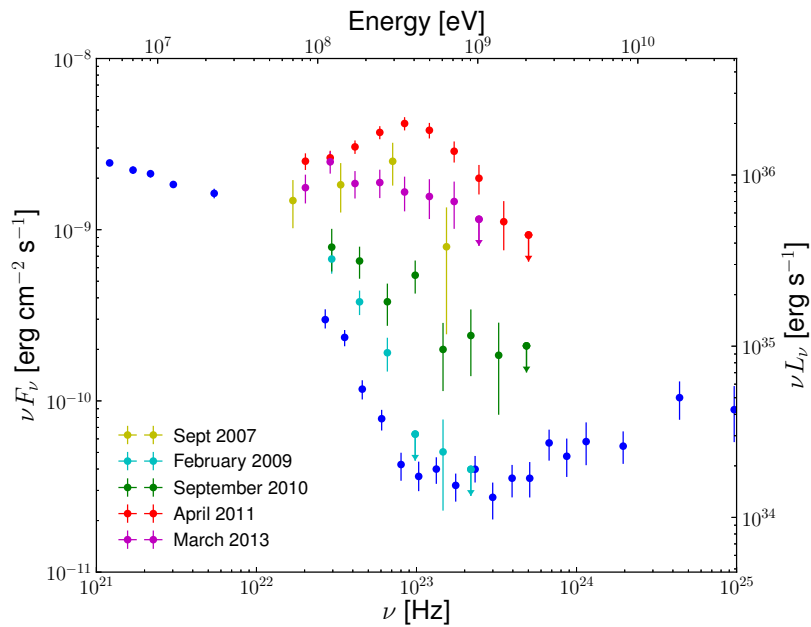}
\caption{Spectral energy distribution at the maximum flux level for five of the six Crab nebula flares detected as of September 2013 \cite{Abdo2011,Buehler2012,Striani2013,Mayer2013}. No spectrum has been published for the low intensity flare of July 2012 \cite{Ojha2012}. The blue points show the average nebula flux values referenced in Fig. \ref{fig:broadsed}. }
\label{fig:flaresed}
\end{figure}

As of September 2013, six flares have been reported  \cite{Buehler2012,Ojha2012,Striani2013,Mayer2013}. Due to the statistical nature of the flux variations, the definition of a flare is somewhat arbitrary. In all flares reported to date the synchrotron component of the nebula had a peak flux $F_{\mathrm{ns,100}} > 35 \times 10^{-7}$~cm$^{-2}$~s$^{-1}$. Equally, the definition of a flare duration is not straight forward; typically, the flux is increased with respect to the monthly average flux for $\sim 1$~week. The SED around the peak of the flares is shown in Fig. \ref{fig:flaresed}. One can see that the spectral behavior differs strongly between flares. During the flare in February 2009, e.g. the flux increased with no measurable spectral changes with respect to the average nebula flux, while during the flare in April 2011 a new spectrum component with a rising flux was observed. Generally, all spectra show significant emission up to $\sim 1$~GeV.

\begin{figure*}[t]
\centering
\noindent\includegraphics[width=0.84\textwidth]{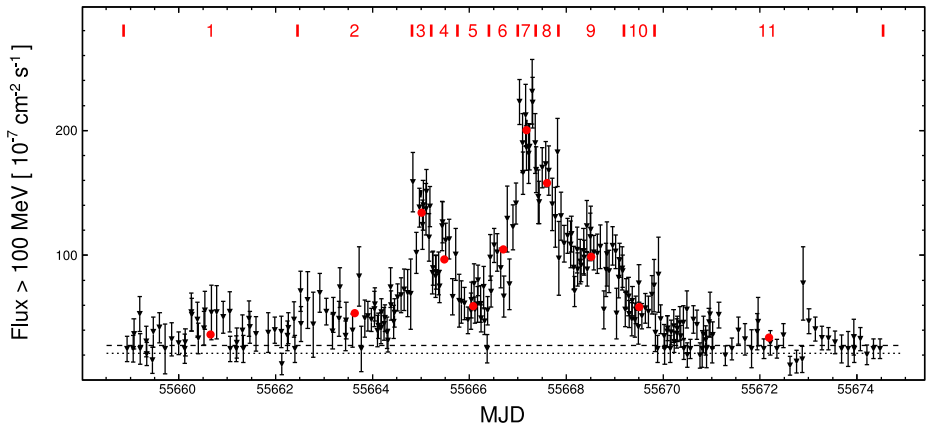}
\caption{Integral flux above 100 MeV from the direction of the Crab as a function of time during the 2011 April flare, reproduced from \citeasnoun{Buehler2012}. The points represent the sum of the nebula and pulsar fluxes. The dotted line indicates the sum of the 33-month average fluxes from the inverse-Compton nebula and the pulsar, which are stable over time. The dashed line shows the flux of the average synchrotron nebula summed to the latter. The red vertical lines indicate time intervals where the flux remains constant within statistical uncertainties. The time windows are enumerated at the top of the panel. The corresponding flux is shown by the red marker below each number. The SED for each of the time windows is shown in Figure \ref{fig:specevol}. }
\label{fig:lcflare}
\end{figure*}

\begin{figure*}[t]
\centering
\noindent\includegraphics[width=0.96\textwidth]{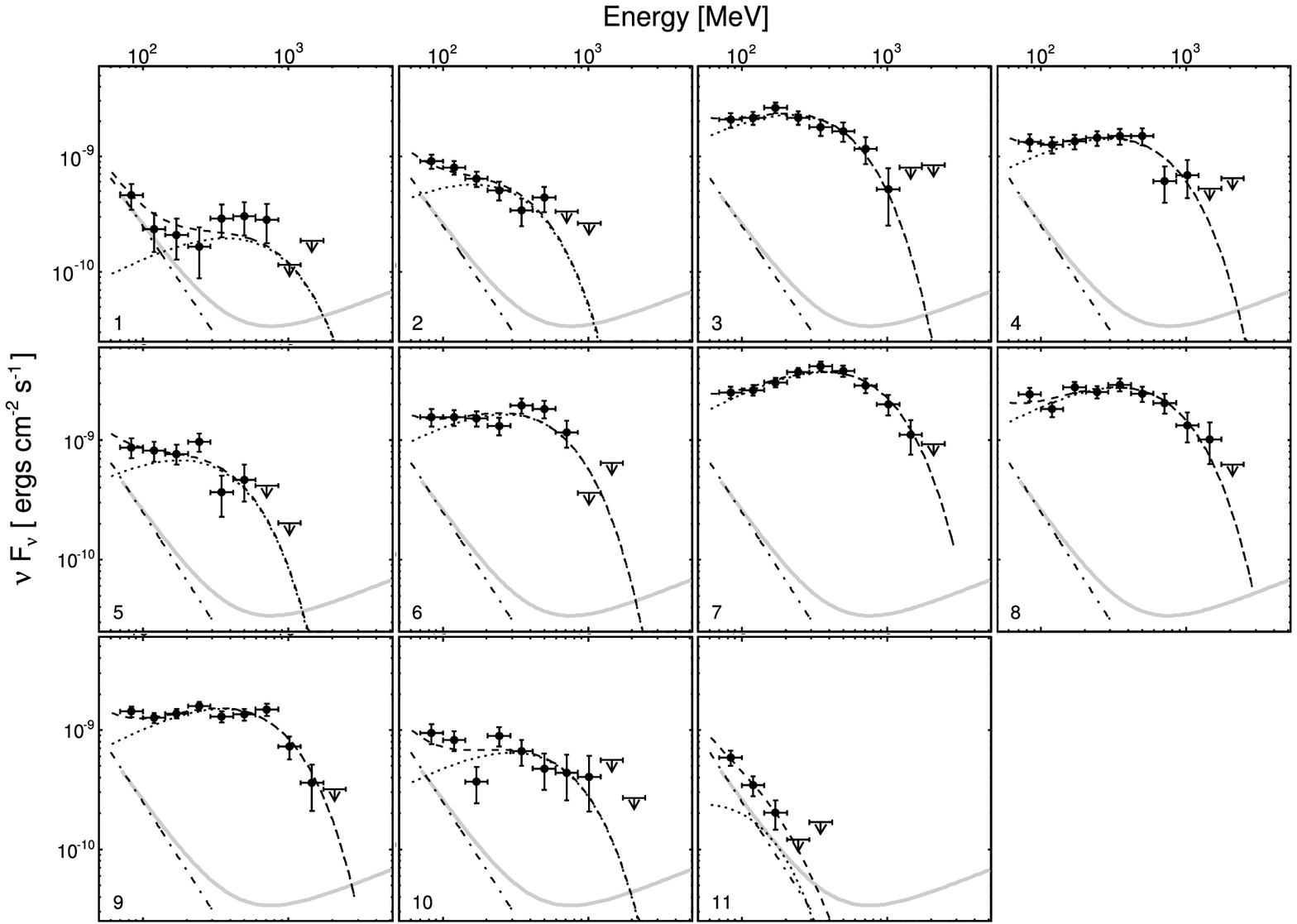}
\caption{Spectral Energy Distribution evolution during the April 2011 Crab flare, reproduced from \citeasnoun{Buehler2012}. The time windows are indicated in the bottom left corner of each panel and correspond to the ones indicated in Figure \ref{fig:lcflare}. The dotted line shows the SED of the flaring component, the dot-dashed line the constant background from the synchrotron nebula, and the dashed line is the sum of both components. The average Crab nebular spectrum in the first 33 months of \textit{Fermi} observations is also shown in gray. }
\label{fig:specevol}
\end{figure*}

The flare of April 2011 gave us the most detailed look into the flare phenomenon to date \cite{Striani2011,Buehler2012}. The light curve of this flare is shown in Fig \ref{fig:lcflare}. Its high flux allowed flux measurements down to time scales of $\sim 20$~minutes. The flux doubled within $t_d \lesssim 8$~hours at the rising edges of the two main bursts during the flare. The spectral evolution is shown in Fig. \ref{fig:specevol}. Interestingly, the apparently complex evolution can be parametrized in a simple way: the emerging spectral component is well characterized by a power-law spectrum with an exponential cutoff. The spectral index $\gamma = 1.27 \pm 0.12$ remains constant within errors during the flare, whereas the cutoff energy $E_\mathrm{C}$ and the total energy flux of the synchrotron component above 100 MeV vary as $L_\mathrm{ns,100} \propto E_\mathrm{C}^{3.42 \pm 0.86}$. At the maximum of the flare, the cutoff energy is $\epsilon_{\mathrm{C,max}} = 375 \pm 26$~MeV and the total isotropic luminosity in the synchrotron component is $L_{\mathrm{max,100}} \approx 4 \times 10^{36}$~erg~s$^{-1}$, approximately 1\% of the spin-down power of the pulsar.  

The angular resolution of current HE instruments is $> 18$ arc minutes, not enough to determine the emission region of the flares within the nebula. From the beginning, it was clear that in order to pinpoint the emission region, correlated variability at radio, optical or X-rays is needed, making use of the $<1$~arc~second angular resolution achieved in these wavebands. However, to date, despite extensive efforts a detection of the flares outside of the HE gamma-ray band remains elusive. Strictly simultaneous observations were obtained during the September 2010, April 2011 and March 2013 flares \cite{Vittorini2011,Morii2011,Lobanov2011,Weisskopf2013,Aliu2014}. Particularly dense observations with the Chandra and Hubble Space Telescopes and the Keck and VLA Observatories were carried out during the April 2011 flare. No increased emission was detected from radio to X-rays for any structure of the nebula above the usual levels. This finding was very unexpected. The inferred flux upper limits show that the SED of the HE gamma-ray flare steeply drops with decreasing frequency, as was already suggested by the hard spectral index of the flaring component in HE gamma-rays during the flare.  

No pulsations are found in the gamma-ray emission of the flares. The pulsar properties remain unchanged during the outbursts, no change in the spin-down period or flux were found in radio, X-rays or gamma-rays \cite{Abdo2011,Morii2011,Buehler2012}. The time scale of the recurrence of pulsar glitches is similar to the recurrence of the gamma-ray flares, however, there is no obvious correlation in time between these two events.

\subsection{The origin of the gamma-ray flares?}

To date, the Crab flares remain mysterious. We do not know what causes them and where they come from within the nebula. Several ideas have been proposed, but no definite answers can be given today. Any explanation will have to encompass \emph{all} the presented observations, so far theoretical models have addressed different aspects of the problem. The rapid variability implies that, unless there is ultrarelativistic beaming, the flare emission comes from a small region within the nebula $R_{\mathrm{f}} \lesssim c \cdot t_d \approx 10^{-4}$~pc, small even when compared to nebula sub-structures as the Sprite, wisps or the inner knot with projected scales greater $10^{-2}$~pc.

A puzzling observation is that flare emission is detected up to photon energies of $\approx 1$~GeV. Synchrotron emission appears to be the only radiation process which is efficient enough to account for the flare emission in the nebula environment \cite{Abdo2011}. However, particles accelerated in MHD flows can only emit synchrotron emission up to a maximum energy $\epsilon_{\mathrm{max}} = 160$~MeV \cite{Guilbert1983,Uzdensky2011}. Therefore, either MHD conditions are not valid in the flaring region, or the emission is relativistically boosted towards our line of sight \footnote{Another alternative is that the flare emission is ``jitter radiation'', which could e.g. be emitted in the striped wind if the length scale of magnetic turbulence is much smaller than the distance between stripes \cite{Teraki2013}}. Both scenarios are possibly interrelated: a breakdown of the MHD conditions occurs in magnetic reconnection events and beaming of particles naturally occurs in the reconnection layer \cite{Zweibel2009,Uzdensky2011,Cerutti2012,Sturrock2012}.

That magnetic reconnection process is responsible for the particle acceleration is also indicated by the fact that the proposed alternatives have severe difficulties: diffusive shock acceleration generally does not produce electron spectra which are hard enough to explain the emission during the April 2011 flare and is expected to be inefficient at the termination of the pulsar wind \cite{Ellison2004,Sironi2011,Sironi2013}, and the proposed acceleration due to absorption of ion cyclotron waves is expected to act on longer time scales \cite{Amato2006}. Magnetic reconnection was studied with PIC simulation in the context of the flares by \citeasnoun{Cerutti2013} and \citeasnoun{Baty2013}. Cerutti et al. show that emission beyond $\epsilon_{\mathrm{max}}$ is possible in such events. The variations in motion of the accelerated particle beams can explain the observed flux variations with a PDS compatible  with the measured one. The spectrum of the April 2011 flare can be qualitatively reproduced, including its dynamical evolution and the observed correlation between cutoff energy and luminosity. The gamma-ray flares are therefore likely connected to explosive reconnection events triggered by current instabilities.

Regions with increased magnetic dissipation are the preferred emission sites, as e.g. downstream of the wind termination, near the inner ring \cite{Bednarek2011,Lyutikov2012,Bykov2012}. In particular, at high latitudes the flow is expected to remain relativistic in the downstream and can be beamed into our line of sight. This produces increased emission from the so called ``arch shock'' of the wind termination, which was associated to the inner knot \cite{Komissarov2011}. Another possibility is that the flares originate from the base of the jet. As pointed out by \citeasnoun{Lyubarsky2012}, turbulence and increased magnetic dissipation are expected in this region, which might be associated to the Sprite. However, no observational signatures related to the flares have been found for any of these regions to date.

The location of the emission region might in principle be different between flares. A distribution of flares throughout the nebula has been proposed, where the beamed gamma-ray emission from different regions randomly crosses our line of sight \cite{Yuan2011,Clausen-Brown2012}. Such a process can produce PDS of flux variations compatible with the observed one, linking the flare phenomenon to the flux variations on longer time scales. The flares might therefore be responsible for a significant part of the magnetic dissipation in the nebula \cite{Komissarov2012}. 

\section{Outlook}

A review of the observations and theory of the Crab is bound to be outdated as it is written. New insights are gained continuously and building up on a long history of research. Nevertheless, key questions for our understanding remain unanswered. In particular, the structure of the magnetosphere, and related to it, the launch of the pulsar wind, remain to be understood in a quantitative way. The observed pulse profiles are sensitive probes of the magnetic field structure and of the sites particle acceleration in the magnetosphere \cite{Bai2010a}. Pulsar emission models can be compared to a wealth of observational data, including phase profiles, spectra and polarization measurements of the pulsar population. The latter is continuously growing, in particular through the ongoing observations of the Fermi-LAT. At the same time, emission models can take advantage of the increasingly more realistic  simulations of pulsar magnetospheres.

Such a broad modeling approach is currently being pursued by several groups, and bears large potential to increase our understanding of pulsars \cite{Bai2010,Romani2010,Harding2011,Kalapotharakos2012}. The VHE gamma-ray pulsations of the Crab are a new constraint which needs to be addressed within this context.  Experimentally, it remains to be seen whether VHE emission is typical for pulsars or only found in the Crab \cite{Collaboration2013a}. Searches by current VHE instruments for pulsations from other pulsars are ongoing. The future Cherenkov Telescope Array (CTA) will allow to perform these searches at higher sensitivities \cite{Hinton2013}.

The understanding of the Crab synchrotron nebula has significantly increased with the help of MHD simulations in the last years. Quantitative comparisons of observations to these simulations can constrain the magnetization and angular dependence of the pulsar wind. First three dimensional simulations show that significant magnetic dissipation is expected to happen downstream of the wind termination \cite{Porth2013}. The gamma-ray flares are an exciting discovery in this context, likely giving us a direct view into magnetic dissipation regions.

It is puzzling that flares have not been detected from any other pulsar wind nebula besides the Crab. The search of flares from other nebulae is ongoing in the HE-gamma-ray band \cite{Ackermann2013}. Giving the unusually large magnetization of the Crab nebula, flare might be found at lower frequencies in other systems. In this case, all-sky X-ray instruments as MAXI are best suited to search for flares from other nebulae \cite{Matsuoka2009}. The HE gamma-ray flares from the Crab have been recurring approximately once per year. Dense multi-waveband coverage in the radio, optical, X-ray and VHE gamma-ray bands was achieved during the last flare in March 2013 \cite{Mayer2013}. The analysis of these observations is still ongoing and might already reveal correlated signatures to the HE-gamma-ray variability. Monitoring programs are already put in place for future flares. The continuous multi-wavelength coverage will enable more sensitive correlation studies and hopefully pinpoint the emission site in the coming years. Also, the detection of an inverse Compton component of the flares at VHE gamma-rays might be possible with CTA in the future, if the emission region is ultra-relativistically beamed towards our line of sight (Lorentz factor $\gtrsim 30$, \citeasnoun{Kohri2012}).

In a broader context, several phenomena observed in the Crab are ubiquitous in the non-thermal Universe. A better understanding of the Crab, and pulsar wind nebulae in general, is therefore interweaved  with our understanding of other sources as active galaxies, microquasars, gamma-ray bursts or novae. The release of magnetic energy by a compact object, and the transfer of this energy into the acceleration of particles is common in these systems. Formation of jets is also ubiquitous in non-thermal sources. This process can be studied in detail in the synchrotron nebula of the Crab. As e.g., the study of the dynamics and formation of the wisps, the jet formation in the Crab allows us to study the behavior of relativistic magnetized plasma and the instabilities that govern its flow \cite{Mignone2013}. The remainders of the ``guest star'' observed almost one thousand years ago will continue to be our companion in these endeavors.

\ack 
We would like to thank all those who have shared our excitement in studying the Crab over the years. We would in particular like to thank Lise Escande, Stefan Klepser, Michael Mayer and Gianluca Giavitto for comments on earlier versions of this manuscript. Rolf Buehler would like to thank Benoit Cerutti, Jonathan Arons, Stefan Funk, Tune Kamae and Lorenzo Sironi for interesting discussions. Finally, we thank Manuel Meyer and Dieter Horns for providing spectral data in tabulated form.

\section*{References}
\bibliography{crabreview}

\end{document}